\documentclass[11pt]{article}
 \usepackage{graphicx}
 \usepackage[labelformat=empty]{caption}
\begin{document}

\noindent{\scriptsize \em
1st. June 2017}

\noindent{\large{\bf  Published tunneling results of Binnig et al interpreted as
related to surface superconductivity in SrTiO$_{3}$}}

\vskip 2em

\noindent
{\bf D.M. Eagles$^1$}

\noindent
Short title: Tunneling and surface superconductivitiy in SrTiO$_3$


\vskip 2em
\noindent {\bf Abstract}

\noindent
In 1980 Binnig et al. reported tunneling measurements on Nb-doped
SrTiO$_3$, and interpreted their results as indicating two-band
superconductivity in the bulk of SrTiO$_3$.  However: (1) Effective
masses determined from tunneling results in the normal state by Sroubek
in 1969 and 1970 are much smaller than those determined by most other
methods. The much smaller masses were attributed to properties of
a surface layer by the present author in 1971;  (2) The only other
reports of two-band superconductivity in bulk SrTiO$_3$ can be used to
infer much smaller values for the band separations than found by Binnig
et al.. In this paper we give an alternative explanation of the results
of Binnig et al. in terms of superconductivity in a surface layer.
We obtain fair fits to the band gaps versus Fermi energy for the two
bands in the three samples where two surface subbands are occupied
and to the temperature dependence of the gaps in one crystal,
using a model with three adjustable interaction parameters, an adjustable
energy for the phonons which dominate the pairing, and an adjustable
ratio of the mean-field $T_c$ to the actual $T_c$.  We show results for a
combined fit to the low-temperature band gaps and to the $T$-dependence
in one crystal.  The phonon energy which gives the best fit is 21 meV.
This is probably an appropriate average over the three longitudinal polar
modes and acoustic modes in the material.  A large value of about two is
found for the ratio $T_{cmf}/T_c$, and we conjecture that this arises
because a band with a small Fermi energy, not seen in the tunneling
results, plays a part in increasing $T_{cmf}/T_c$.

\noindent
$^1$ 19 Holt Road, Harold Hill, Romford, Essex RM3 8PN, England 


\newpage
\twocolumn
\noindent
{\bf 1. Introduction} 

\vskip 1em
\noindent
In 1980 Binnig et al. \cite{Bi80} claimed that tunneling results into
(100) surfaces of SrTiO$_3$ reported by them gave evidence for two-band
superconductivity in bulk Nb-doped SrTiO$_3$.  However  
I have had doubts for some time as to whether what they were observing
was bulk superconductivity.  Two reasons for thinking this are: (1)
Effective masses determined from tunneling results of Sroubek in the
normal state \cite{Sr69,Sr70} are much smaller than those determined by
most other methods. The much smaller masses were attributed to properties
of a surface layer by the present author in 1971 \cite{Ea71}.
(2) There are two other reports of two-band superconductivity in bulk
SrTiO$_3 $\cite{Li14a,Li14b}, but one of these \cite{Li14a} can be
shown to imply much smaller values for the band separation than the
32 meV reported in \cite{Bi80}. This may be seen from the statement
in \cite{Li14a} that the Fermi energy when the second band starts to
become occupied is 1.8 meV.  A third band, which may or may not play a
part in the superconductivity, starts to become occupied at a carrier
concentration about nineteen times higher, corresponding to a Fermi energy
of the order of $19^{\frac{2}{3}} \times$ 1.8 meV, i.e. $\sim$ 13 meV, still
much smaller than the 32 meV mentioned in \cite{Bi80} for the occupation
of the second band.  

In this paper we give an alternative explanation for the results
of \cite{Bi80} where the two gaps observed are associated with the
lowest two subbands of a surface layer.  At most two subbands of the
band which we consider for most of this paper are occupied for the
crystals studied.  We make use of a two-subband model, and use the type
of equations found by Suhl et al. \cite{Su59} to analyse the data,
but we suppose that the mean-field transition temperature $T_{cmf}$
may be larger than the observed $T_c$ where the energy gaps vanish.
Although the electron-phonon coupling in the normal state is largest for
the highest-energy longitudinal optical phonon \cite{Ea65,Ba66,De10},
this may not be the case for the superconducting state because  phonons
having a higher energy than the Fermi energy may have reduced effects on
the superconducting pairing \cite{Sw16,Ro16}.  Hence we use an adjustable
energy for the phonons which dominate the pairing.  A large value of
about two is found for the ratio $T_{cmf}/T_c$.  We conjecture that
this arises because a band with a small Fermi energy, not seen in the
tunneling results, plays a part in increasing $T_{cmf}/T_c$.

Fernandes et al. \cite{Fe13} discuss two-band superconductivity in bulk
SrTiO$_3$, but use a two-dimensional model for much of their analysis.
They also note that Binnig et al.'s tunneling results are sensitive
to surface states, and point out that three-dimensional theory
does not present a kink in the energy gap which is observed when the
second band starts to become occupied.  They further find that interband
pairing is small compared with intraband pairing, and so their results
have some similarity to ours.   However, they differ in several ways.
First they do not claim as we do, that the results of \cite{Bi80} are
entirely associated with surface effects.  Secondly, they do not give
any detailed fits to results as we do.  Thirdly they give a different
reason from us for the dominance of intraband over interband pairing.
Fourthly they assume an upper cut-off energy for the interaction 
of 160 meV, higher than the energy of the highest energy phonon,
whereas our fitting with an adjustable phonon energy gives a much smaller
cut off of about 21 meV.

Another theoretical paper somewhat related to ours is a discussion of
multigap structure in electric-field-induced surface superconductivity
by Mizohata et al. \cite{Mi13}.  Values of surface carrier density and
electric field are chosen to be relevant to experiments on SrTiO$_3$
by Ueno et al. \cite{Ue08}.

Our use of a model where interaction with longitudinal optical phonons
dominates the pairing has some support from a paper by Klimin et
al. \cite{Kl17}, who, using a dielectric approach to pairing, find that
such phonons dominate the pairing.  However, from their paper it is not
clear to me the relative importance of the three types of longitudinal
polar optical phonons in their theory.  Transport measurements \cite{Fr67,
Ma11}, ARPES in SrTiO$_3$ \cite{Wa16} and on LaAlO$_3$-SrTiO$_3$
bilayers \cite{Bo15}, and tunneling \cite{Sw16}, all indicate that most
interactions in the normal state, at least at low carrier concentrations,
are with the two highest frequency longitudinal optical phonons,
as expected from theory.  On the other hand,
in the superconducting state Swartz et al. \cite{Sw16} find much weaker
coupling, which they attribute to the fact that the energy of the high
energy phonons is larger than the Fermi energy in superconducting
crystals.  They conjecture that low energy phonon modes may be most
important for the superconductivity, but to me it appears possible that
the high energy phonons may still be important, but with interactions
giving rise to pairing relatively small because of the low $E_F$.
Baratoff and Binnig \cite{Ba81} argue that the highest frequency
longitudinal optical phonon is important for superconductivity because
of the sharp fall in $T_c$ when the plasmon frequency becomes greater
than the frequency of this mode.

In section 2 we analyse the energy gaps at low temperatures as a
function of Fermi energies, and in section 3 we analyse the temperature
dependence of the energy gaps reported in \cite{Bi80} in one crystal.
We have three coupling parameters and the phonon energy relevant for the
superconductivity for our low-temperature fitting, and one extra parameter,
$T_{cmf}/T_c$, for the fitting of the temperature dependence of the
gaps in one crystal.  In order not to waste data, in section 3 we find a
combined fit for the energy gaps at low temperatures in three crystals
showing two gaps, and for the temperature dependence of the energy gaps
in one crystal, with appropriate weightings for the two types of data.

The phonon energy which gives our best fit is about 21 meV, but another
minimum of our sum of squares between theory and experiment at only a
slightly larger value occurs at about at about 28 meV, with a not very
high saddle point in between at about 25 meV. Either the 21 meV or 28
meV minimum is probably some average of the three polar longitudinal
phonon energies and perhaps acoustic modes, interaction with which
was introduced empirically in \cite{Kl17}.  Although the strongest
coupling to the phonons in the normal state is to the highest frequency
longitudinal phonon of energy 99 meV \cite{Ea65}, in the superconducting
state there is a reduction of its contribution to the pairing because
its energy is greater than the Fermi energy for the crystals being
considered (see e.g.\cite{Sw16,Ro16,Ru16}), and also, for pairing,
there is a further reduction relative to that with the lower energy
phonons because the phonon energy appears in an energy denominator for
the phonon-induced pairing.  Although the coupling to the lowest energy
longitudinal polar phonon is negligible at low carrier concentrations 
even after a numerical error giving the coupling a factor of the order
of ten too small in \cite{Ea65} is corrected (cf \cite{Ba66}), when
finite carrier concentrations are considered \cite{Me10} this coupling
may be non-negligible.

In bulk SrTiO$_3$ there is evidence from the isotope effect \cite{St16}
that the ferroelectric soft mode plays an important part in the pairing,
as predicted in \cite{Ed15,Ke16}, but in surface layers, because of the
lower static dielectric constant due to large electric fields, this mode
is less likely to be of importance.

The ratio of intersubband to intrasubband interactions depends on the
range of interactions giving rise to pairing in comparison with the width
of the surface layer. To calculate this ratio for a realistic form of
interactions would be quite difficult. However,  in section 4, we  consider
a one-dimensional square-well model for the interaction potential,
and show how, in this model, the ratio of intersubband to intrasubband
interactions depends on the ratio of the width $w$ of the square well to
the width $d$ of the surface layer.  We also find the value of $w/d\approx
0.47$ which gives a value for this ratio in agreement with the value of
about 0.14  determined from our data fitting.  This value of 0.14 is
smaller than the value of 2/3 found for local interactions \cite{Pa65}.
Paskin and Singh's result that the intrasubband interaction is increased
by a factor of 3/2 for local interactions for very thin films was argued
by us to be increased by a factor 9/4 in very thin whiskers, and was used
speculatively in combination with enhanced densities of states at low
carrier concentrations in such films or whiskers to estimate high
transition temperatures in (111) films and [111] whiskers of 0.7, 1.4
and 2.1 nm thicknesses in \cite{Ea67}.

\vskip 2em
\noindent
{\bf 2. Energy gaps as a function of Fermi energy} 

\vskip 1em
\noindent
Tunneling into eight Nb-doped crystals was reported in
\cite{Bi80}. The Fermi energies for these crystals varied from about
18 to 60 meV.  From results on three of the crystals it was found
that the second band commenced in energy at about 32 meV above the
bottom of the lowest band.  In our interpretation we regard the two
bands as surface subbands.  If we were to assume wave functions which
vanish at the edges of a surface layer of given thickness, then the
third subband would commence at an energy of about $(8/3)\times 32$
meV$\approx 85$ meV. However, by studying the type of model for the
surface layer used in \cite{Ea71} we see that the barrier at the edge
of the surface layer does not extend above the Fermi energy, and so,
for a maximum Fermi energy of about 60 meV in the crystals studied in
\cite{Bi80}, there will be no third surface subband.  We also ignore
any possible pairing interaction between bulk and surface subbands.
Although the fact that the superconducting $T_c$ drops sharply when
the plasmon energy exceeds the energy of 99 meV of the highest energy
longitudinal optical phonon has been argued to imply that that phonon
dominates the pairing \cite{Ba81}, because of the generally held view
that pairing mediated by phonons of energy larger than or of the order
of the Fermi energy is smaller than it would be for large Fermi energies
(see e.g. \cite{Sw16,Ro16,Ru16}), we allow the phonon energy involved
in the pairing to be adjustable, and assume that any value we find is
an appropriate average over several phonon energies.  The limits to the
integrals involved in the BCS-type of theory have to be altered if the
Fermi energy measured from the bottom of any band is smaller than the
phonon energy.

The densities of states of the subbands would all be equal within the
subbands if the surface layer were of a constant thickness.  However,
since the layer thickness will increase  as the energy increases, we allow
for differing densities of states in the two subbands.  An important
parameter appearing in our theory is the ratio $r$ of the pairing
potential between subbands to that within the lowest subband.  Although,
for local pairing interactions this ratio is 2/3 \cite{Pa65}, we shall
allow this ratio to be adjustable to be determined from our data fitting.
In section 4 we shall show that, for a square-well type of pairing potential
$r$ beomes smaller when the size of the square well becomes comparable
with the width of the surface layer, and that an empirical value of about
0.14 determined by our data fitting to two types of data in section 3 will
occur for a square-well type of pairing potential if the ratio $w/d$ of
the size $w$ of the square well to the layer thickness $d$ is 0.47.

For a two-band superconductor with gap parameters $\Delta_i, i=1,2$,
with constant densities of states in each subband between the bottoms
of the bands and $E_{Fi}+\hbar\omega$, at $T=0$ there are two quantities
$F_i(\Delta_i)$ ($i$=1,2) analogous to $F(A)$ on page 553 of \cite{Su59}, given by
\begin{eqnarray}
F_i(\Delta_i)=(1/2)\int_{-p_i}^{\hbar\omega}[1/(\epsilon^2+\Delta_i^2)
^{\frac{1}{2}}\nonumber\\=(1/2)[\rm{arcsinh}(p_i/\Delta_i)+\rm{arcsinh}
(\hbar\omega/\Delta_i)],
\end{eqnarray}
where $p_1=min(|E_{F}|,\hbar\omega)$, and
$p_2=min(|E_F-E_c|,\hbar\omega)$, where $E_F$ is the Fermi energy
measured from the bottom of the lower band, and $E_c$ is the energy of
the bottom of the upper band above that of the lower band.  We have made
use of the fact that the tanh in the expression for $F(A)$ in \cite{Su59}
is replaced by unity at $T=0$.

Provided $p_i$ and $\hbar\omega$ are significantly greater than
$|\Delta_i|$, then the arcsinh's in equations (1) can be approximated by
the logarithms of 2$p_i/\Delta_i$. 

We have four parameters to consider in this section.  These may be taken
to be $\lambda_1$ and $\lambda_2$, the densities-of-states interaction
products within the two subbands, the ratio $r$ of intersubband
interactions to intrasubband interactions, and the phonon energy
$\hbar\omega$. 

When two bands are occupied, equation (4) of \cite{Su59} gives, at
$T=0$, using our notation introduced above, the following two simultaneous
equations to solve:
\begin{equation}
\Delta_1=\Delta_1\lambda_1 F_1(\Delta_1)+r\Delta_2\lambda_2 F_2(\Delta_2),
\end{equation}
\begin{equation}
\Delta_2= \Delta_2\lambda_2 F_2(\Delta_2)+r\Delta_1\lambda_1 F_1(\Delta_1).
\end{equation}
If only one band is occupied then equation (3) does not occur, and 
the second term on the right-hand side of equation (2) 
is omitted. 

For the crystals with two subbands partially occupied, we treat the three
coupling parameters as adjustable.  We solve equations (2) and (3) numerically
for crystals 6, 7 and 8, and then, we find the values of $\lambda_1,
\lambda_2$ , $r$ and $\hbar\omega$ which give the best root-mean-square
difference between calculated and six observed energy gaps for crystals
6, 7 and 8, after making small estimated corrections due to the non-zero
value of temperature at which the energy gaps were found.  However, with
four parameters and only six data points, uncertainties in parameters
will be too large for their values to have much meaning, and so we do
not present results for parameters until a combined fit to two types of
data for a total of 29 points has been performed in the next section.

\vskip 2em
\noindent
{\bf 3. Temperature dependence of gaps in one crystal} 

\vskip 1em
\noindent
In the work mentioned in the last section, we were able to find a fair
fit to the energy gaps as a function of Fermi energy determined by
Binnig at al. \cite{Bi80} using a model where the superconductivity
is associated with two surface subbands, although we postponed showing
results there because of insufficient data to determine parameters well.
In this section we examine the temperature dependence of the energy gap
in one crystal using the same type of model.  The transition temperature
for the crystal considered may be estimated to satisfy $k_BT_c \approx$
54 $\mu$eV from figure 2 of \cite{Bi80}.

The equations we use are:
\begin{equation}
\Delta_2\lambda_{12}s_2=
\Delta_1(1-\lambda_1 s_1)
\end{equation}
and
\begin{equation}
\Delta_1\lambda_{21}s_1=
\Delta_2(1-\lambda_2 s_2).
\end{equation}
Here $\Delta_1$ and $\Delta_2$  are the two energy-gap
parameters,  $\lambda_i, i=1,2$, are the intrasuband values of the
density-of-states interaction products, 
$\lambda_{21}=\lambda_1r,
\lambda_{12}=\lambda_2r$, where $r$ is the ratio of intersubband
to intrasubband interactions. 
Also
\begin{equation}
s_1=0.5s_{11}+0.5s_{12},
\end{equation}
and
\begin{equation}
s_2=0.5s_{21}+0.5s_{22},
\end{equation}
where
\begin{equation}
s_{11}=\int_0^{q_1} f_1(x)dx,
\end{equation}
\begin{equation}
s_{12}=\int_{0}^{\hbar\omega} f_1(x)dx,
\end{equation}
\begin{equation}
s_{21}=\int_0^{q_2} f_2(x)dx,
\end{equation}
\begin{equation}
s_{22}=\int_{0}^{\hbar\omega} f_2(x)dx.
\end{equation} 
Here $q_1=min(59.5 {\rm mev},\hbar\omega)$, $q_2=min(59.5 {\rm meV}-E_c, \hbar\omega)$ and
\begin{equation}
f_1={\rm
tanh}[(x^2+\Delta_1^2)^{\frac{1}{2}}/2t]/(x^2+\Delta_1^2)^{\frac{1}{2}},
\end{equation}
\begin{equation}
f_2={\rm
tanh}[(x^2+\Delta_2^2)^{\frac{1}{2}}/2t]/(x^2+\Delta_2^2)^{\frac{1}{2}},
\end{equation}
with $t=k_BT_{cmf}(T/T_c)$, where $T_{cmf}$ is the mean-field
transition temperature,  $T_c$ is the temperature where the observed gap
vanishes, and $E_c$ is the energy of the bottom of the second subband
above the bottom of the first. 

Assuming $E_c=32$ meV as found in \cite{Bi80}, for given $\lambda_1,
\lambda_2, r, \hbar\omega$ and $T_{cmf}$, we have two simultaneous
integral equations to solve to obtain the energy gaps for a given
temperature.  We treat this problem as a subroutine, and then try
minimising the sum of squares of differences between theoretical and
experimental reduced energy gaps as a function of $T$, where experimental
points for both energy gaps are shown in the inset of figure 2 of
\cite{Bi80}.  If we take $k_BT_c$ of this crystal to have the value
which can be inferred from fugure 1 of \cite{Bi80}, i.e. $k_BT_c\approx
0.66\Delta_{8}(0)$, where $\Delta_{8}(0)$ is the larger energy gap at
$T=0$ for this crystal inferred from the observed gap at $T\approx 0.2
T_c$, we find that we cannot obtain a good fit for all temperatures.
So we adopt a different strategy, supposing that the $T_c$ inferred
from  figure 2 of \cite{Bi80} is smaller than the mean-field transition
temperature $T_{cmf}$ because in general mean-field theory overestimates
$T_c$.  So as not to waste data, we perform a combined fit to the results
for the energy gaps at low temperatures for the crystals with two energy
gaps, as discussed in section 2, and to the temperature dependence of
the gaps in one crystal, with weightings for the two types of points
inversely proportional to the squares of the average values for the two
types of points in the units used.

The best fit to a total of 29 points, six for low-$T$ energy gaps in
three crystals, and 23 for the $T$-dependence of the energy gap in one
crystal, gives a root-mean-square error of 5.5\% of the mean values
of the experimental points.  The values of parameters for the fit are
$\lambda_1=0.1478, \lambda_2=0.1181, r=0.143, \hbar\omega=21$ meV and
$T_{cmf}/T_c$=2.055.  The theoretical and experimental data for the above
parameters are shown in figures 1 and 2.  
\begin{figure}
{\centerline{\includegraphics[width=7.5cm, angle=270]{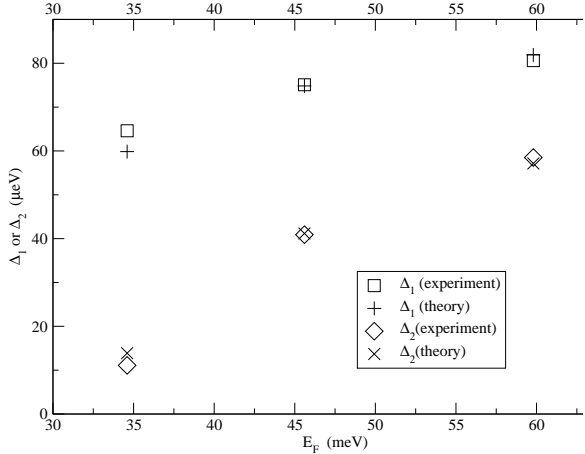}}}
\caption*{FIG. 1.  Comparison of theory and experiment for energy gaps at low
$T$ for the three crystals with two gaps. The experimental values are
taken approximately from figure 2 of \cite{Bi80}.} \end{figure}
\begin{figure}
{\centerline{\includegraphics[width=7.5cm, angle=270]{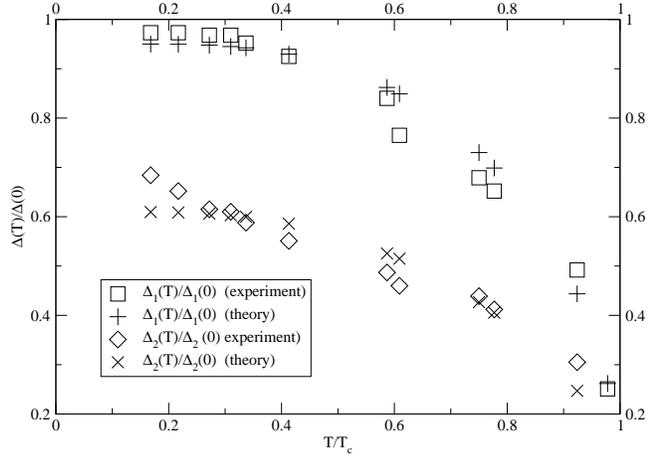}}} \caption*{FIG. 2.  Comparison of theory and experiment for  the temperature dependence of
the energy gaps in one crystal.  The experimental values are taken
approximately from figure 2 of \cite{Bi80}.} \end{figure} We note that
$\lambda_2$ from our fitting is slightly smaller than $\lambda_1$,
as expected, since the effective width of the surface layer will be
larger for a larger mean energy of a subband (see figure 1 of \cite{Ea71}).
The value of $T_{cmf}/T_c$ is required to be greater than unity but with
$T_{cmf}$ being of the same order of magnitude as $T_c$.  Values found
from our data fitting are consistent with these requirements.  Further
section 4 gives some restrictions on the value of the ratio of inter to
intrasubband interactions if we have some information about the range
of the pairing potential to the width of the surface layer.

By taking fixed values for the phonon energy and varying the
other parameters we find that there is quite a large uncertainty
in the phonon energy on the higher energy side, and we estimate
that $\hbar\omega=21^{+8}_{-1}$ meV, with a small region around
$\hbar\omega\approx 25$ meV where the sum of squares of differences
between theory and experiment is larger than the standard error (68\%
confidence level).

For the five crystals with only one subband occupied, the interaction
parameter $\lambda_1$ could have a different value from what it has in
the crystals with two subbands occupied  because of different screening,
and so fitting these points with a different value of $\lambda_1$ will
not help in determination of parameters.  However, if we assume that
$\hbar \omega$ = 21 meV, we find that $\lambda_1$ varies from 0.142 and
0.148 for the five crystals, with an average of 0.144.  These values
are not much different from the value of $\lambda_1=0.1478$ found by fitting
29 points in the this section.

\vskip 2em
\noindent{\bf 4. Estimate of the ratio of intrasubband to intersubband
pairing strengths for a square-well potential}

\vskip 1em
\noindent
In this section, assuming a square-well pairing potential,
we calculate the ratio $r$ of intersubband to intrasubband pairing
strengths as a function of the ratio of the size of the square well in
the direction normal to the surface to the surface thickness, and find
the size of the square well which gives agreement with our empirical
value of  $r$ determined in the last section.

We suppose that we have a surface layer of thickness $d$ with vanishing
wave functions at $z=0$ and $z=d$ , where $z$ is the coordinate
perpendicular to the surface.  We also suppose that, after integrating
over coordinates parallel to the surface, we have a potential well
$V(z-z')$ of constant magnitude between electrons at ${\bf r}$ and ${\bf
r'}$ for $|z-z'|<z_c$ and zero otherwise.  To find the pairing strength
between subbands $n$ and $n'$ we numerically integrate
\begin{eqnarray}
(2/d^2)\int_0^d\int_0^d {\rm sin}(\pi nz/d){\rm sin}(\pi nz'/d)\nonumber\\ V(z-z')
{\rm sin}(\pi n'z/d){\rm sin}(\pi n'z') dz dz',
\end{eqnarray}
and find the ratio of the pairing strength between subbands 1 and 2 to
the intrasubband pairing strength within subband 1 as a function of $w/d$,
where $w$ is the range of the pairing interaction.  For a square-well
type of pairing potential some results are shown in figure 3.  We see
\begin{figure}
{\centerline{\includegraphics[width=7.5cm,]{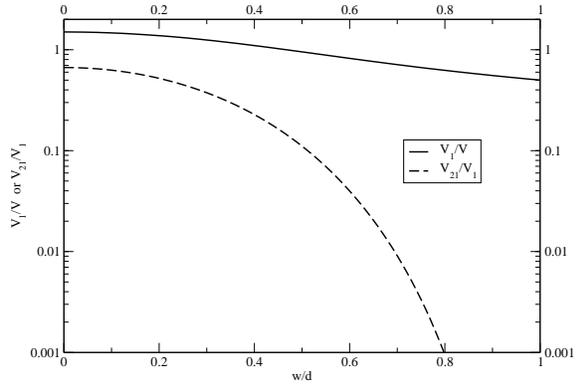}}}
\caption*
{FIG. 3.  Ratio of interband to intraband pairing strengths $V_{21}$ and $V_{1}$
and ratio of $V_1$ to the pairing strength $V$ in the bulk as a function
of $w/d$.}
\end{figure}
that when $w$ becomes comparable with $d$ the ratio $r$ can become small.
When $w/d>0.46$, the intrasubband interaction is decreased below what
it would be for the bulk if we ignore other changes such as
increased densities of states for thin layers.

We see from the figure that $r=V_{21}/V_{11}=0.14$ when the ratio
$w/d\approx$ 0.47.

We suppose that the bare effective mass in the $z$-direction is equal to
a heavy bare mass of perhaps about $6m_e$ (this mass will depend on the
carrier concentration, as can be inferred from figure 3 of \cite {Ma11}),
and in \cite{Ea15} we estimated that it was about 7.3$ m_e$ for a carrier
concentration of 9.4 $\times 10^{20}$ cm$^{-3}$).   With a mass of 6$m_e$
in the $z$-direction, the  energy of the first excited state of about
32 meV from \cite{Bi80} implies that the surface state thickness $d$
is given by
\begin{equation}
3(\pi/d)^2\hbar^2/(2\times 6 m_e)=32\;{\rm meV}.  
\end{equation} 
This gives $d\approx$ 2.4 nm.

Probably a characteristic length for the interaction between two
large polarons will be the radius of a polaron with bare mass equal
to the reduced bare mass of two polarons, i.e. half the bare mass of
a polaron.  If we assume a bare mass of in the $z$-direction of 6m$_e$
as discussed above, then the radius for a particle of half this mass
will be $(\hbar/6m_e \omega)^{\frac{1}{2}}$.  If $\hbar\omega=0.021$ eV,
then this radius is 0.78 nm, giving a diameter of 1.56 nm.  Thus the ratio of
this to the width of the surface region is 1.56/3.4=0.46. This is very
close to the of 0.47 found empirically by us for a square-well porential
and  a surface barrier also of square-well shape.  In view of the fact
that neither a square well for the inter-particle potential  nor the
surface potential are realistic, the good agreement of the estimated
ratio of $w/d$ with the empirically determined ratio must probably be
regarded as a lucky coincidence.

For the interaction between two large polarons at $r_1$ and $r_2$, judging
by the potential due to a single polaron \cite{Fr54}, the interaction
starts flat at zero separation, and then changes to an interaction
of the form proportional to $1/(r_1-r_2){\rm exp}[-(r_1-r_2)/r_{pr}]$, where
$r_{pr}$ is the polaron radius for a particle with bare mass of half the
single-particle mass, as discussed above.  Thus, apart from the tail
to the potential, the interaction is not far from a square-well form.
There may also be a deepening of the interaction potential with a much
shorter range due to deformation-potential interaction with acoustic
phonons.

For the surface potential we expect a sharp rise at the boundary between
the surface region and the bulk, where the effective mass changes, but
a more gradual change of potential near the surface whose detail will
depend on the electric-field dependence of the dieliectric constant, with
a form qualitatively similar to that shown in figure 1 of \cite{Ea71}.

\vskip 2em 
\noindent
{\bf 5.  Remarks on the large ratio of $T_{cmf}$
to $Tc$ found by our data fitting}

\vskip 1em
\noindent
Differences between mean-field transition temperatures and observed
transition temperatures are usually thought to arise because phase
coherence occurs at a lower temperature than the pairing temperature.
In the present problem we are taking the observed transition temperature
as the temperature where the tunneling gap vanishes.  When there is
pairing but no phase coherence of the pairs, there will still be energy
gaps, but these can be expected to spread over a range of values due
to the thermal spread of the center-of-mass energy, and so no sharp peaks
in the tunneling spectrum can be expected except at low temperatures.

The difference between the pairing temperature and the temperature at
which there is condensation of the pairs can arise because of being
close to or in the Bose-Einstein condensation (BEC) region \cite{Ea69},  and can be
enhanced by low dimensionality.  For the crystal for which the temperature
dependence of the gap has been reported, the gap is several orders of
magnitude smaller than the Fermi energy, and so we are not close to
the BEC region.  However, with only two subbands partially occupied,
we are close to two-dimensionality.  We have not made an exhaustive
literature search for differences between the mean-field $T_c$ and
the actual $T_c$, but note that L. Miu \cite{Mi08} analyses results on
resistivity of YBaO$_{6.5}$ and finds that the ratio of the mean-field
T$_c$ to the Berezinsky-Kosterlitz-Thouless temperature is (73/47)=1.55.
This is not quite as large as our estimated value of just over two.  As a
first guess we might conjecture that the lower carrier concentration in
SrTiO$_3$ could contribute to the larger value.   However, for a one-band
model, Kagan finds theoretically \cite{Ka16} that
\begin{equation} 
|T_c^{BCS}-T_c^{BKT}| \sim T_c^{BCS}/E_F, 
\end{equation}
which is small compared to one in our system.   Also, Chubukov et
al. \cite{Ch16} find that, for a two-band model with interband pairing
dominating, if $E_F>>E_0$, where  $2E_0$ is a pair binding energy,
then one also gets $T_c\approx T_{ins}$, where $T_{ins}$ is the onset
temperature of the pairing, presumably to be identified with $T_{cmf}$.
Thus, it seems that we need to bring in further postulates to explain
the large ratio of $T_{cmf}/T_c$.   One possibility is that, beside the
two subbands discussed in the main part of the paper, there is a bulk or
surface band which has a larger gap, and a small carrier concentration
and Fermi energy, for which the transition to the BEC regions is reached
or close to being reached.  Bands with small occupations and Fermi
energies have been discussed by several authors in connection with
underdoped cuprates \cite{Pe00}, some of the Fe-based superconductors
\cite{Lu12}, and in connection with shape resonances associated
with quantum confinement effects \cite{In10,Ma17}.

\vskip 2em
\noindent{\bf 6. Conclusions} 

\vskip 1em
\noindent
It has been argued that Binnig et al.'s tunneling data \cite{Bi80} in
Nb-doped SrTiO$_3$ are associated with superconductivity in a surface
layer.  Fair fits to the energy gaps at low temperature in three crystals
and to the temperature dependence of the energy gap in one crystal
have been obtained using a model with three adjustable interaction
parameters, an adjustable phonon energy for the phonons which dominate
the pairing, and one other partially adjustable parameter, viz. the
ratio of the mean-field transition temperature to the observed $T_c$.
The phonon energy found by data fitting is approximately 21 meV, which
probably represents a weighted energy of the phonons which contribute
to the pairing.  A large value found for $T_{cmf}/T_c$ may imply that a
band with small occupation and Fermi energy plays a part in increasing
$T_{cmf}$.  The ratio $r$ of intersubband to intrasubabnd interactions
from our data fitting is about 0.14, smaller than the value of 2/3 for
local interactions.  We show that, for a square-well pairing potenti,
a value $r=0.14$ is obtained when the ratio of the width of the square
well to the width of the surface layer is 0.47.

\noindent
$^{*}$ E-mail: d.eagles@ic.ac.uk


\noindent
\begin{thebibliography}{99}

\bibitem{Bi80} Binnig, G.,  Baratoff, A.,  Hoenig, H., Bednorz, J.G.: 
Phys. Rev. Lett. {\bf 45}, 1352 (1980)

\bibitem{Sr69} Sroubek, Z.:  Solid State Commun. {\bf 7}, 1561 (1969)

\bibitem{Sr70} Sroubek, Z.: Phys. Rev. B {\bf 2}, 3170 (1979) 

\bibitem{Ea71} Eagles, D.M.: Phys. Stat. Sol. (b) {\bf 48}, 407 (1971)

\bibitem{Li14a} Lin, X., Gourgout, A.,  Seyfarth, G.,  Kr\"amer, S.,
Nardone, Mi., Fauqu\'e, B., Behnia, K.: Phys. Rev. Lett. {\bf 112}, 207002
(2014)

\bibitem{Li14b} Lin, X., Gourgout, A., Bridoux, G., Jomard, F., Pourret, A.,
Fauqu\'e, B., Aoki, D., Behnia, K.: Phys. Rev. B {\bf 90}, 140508 (2014)

\bibitem{Su59} Suhl, H., Matthias, B.T., Walker, I.R.:  Phys. Rev. Lett.
{\bf 3}, 552 (1959)

\bibitem{Ea65} Eagles, D.M.: J. Phys. Chem. Solids {\bf 26}, 672 (1965) 

\bibitem{Ba66} Barker Jr, A.S.: Phys. Rev. {\bf 145}, 291 (1966) 

\bibitem{De10} Devreese, J.T., Klimin, S.N., van Mechelen, J.L.M., 
van der Marel, D.: Phys. Rev. B {\bf 81}, 125119 (2010) 

\bibitem{Sw16} Swartz, A.G., Inoue, H., Merz, T.A., Hikita, Y., Raghu, S.,
Devereaux, T.P., Johnston, S.,  Hwang, H.Y.: Proceedings SPIE {\bf 9931}
(2016) [{\em arXiv}:1608.05621]

\bibitem{Ro16} Rosenstein, B.,  Shapiro, B.Ya., Shapiro, I., Li, D.: 
Phys. Rev. B {\bf 94}, 02405 (2016)

\bibitem{Fe13} Fernandes, R.M., Haraldsen, J.T., W\"olfe, P., Balatsky, A.V.:
Phys. Rev. B {\bf 87}, 014510 (2013) 

\bibitem{Mi13} Mizohata, Y., Ichioka, M., Machida K.: Phys. Rev. B 
{\bf 87}, 14505 (2013) 

\bibitem{Ue08} Ueno, K., Nakamura, S., Shimotani, H., Ohtomo, A., Kimura, N.,
Nojima, T., Aoki, H., Iwasa, Y., Kawasaki, M.: Nat. Mater. {\bf 7}, 855
(2008) 

\bibitem{Kl17} Klimin, S.N., Tempere, J., Devreese, J.T., Van der
Marel, D., J. Sup. Nov. Mag. {\bf 30}, 757 (2017) 

\bibitem{Fr67} Frederikse, H.P.R., Hosler, W.R.: Phys. Rev. {\bf 161},
2 (1967) 

\bibitem{Ma11} Van der Marel D., Van Mechelen, J.L.M., Mazin, I.I.:  
Phys. Rev. B {\bf 24}, 205111 (2011)

\bibitem{Wa16} Wang, Z.,  McKeown Walker, S., Tamai, A., Ristic, Z., 
Bruno, F.Y., De la Torre, A., Ricc\`o, S., Plumb, N.C., Shi, M., Hlawenka, P.,
S\'anchez-Barriga, J., Varykhalov, A., Kim, T.K., Hoesch, M., King, P.D.C.,
Meevasana, W., Diebold, U., Moritz, B., Devereux, T.P., Radovic, M.,
Baumberger, F.: Nature Mat. {\bf 15}, 835 (2016)

\bibitem{Bo15} Boschker, H., Richter, C., Fillis-Tsikaris, E., Schneider, C.W., 
Mannhart, J.: Scientific Reports {\bf 5}, 12309 (2015) 

\bibitem{Ba81} Baratoff, A., Binnig, G.:  Physica {\bf 108B}, 1335 
(1981)

\bibitem{Ru16} Ruhman, J., Lee, P.A.: Phys. Rev. B {\bf 94}, 224515 (2016)

\bibitem{Me10}  Meevasana, W., Zhou, X.J., Moritz, B., Chen, C.-C.,
He, R.H., Fujimori, S.-I., Lu, D.H., Mo, S.-K., Moore, R.G., Baumberger, F.,
Devereaux, T.P.,  Van der Marel, D., Nagaosa, N., Zaanen, J., Shen, Z.-X.:
New Journal of Physics {\bf 12} 023004 (2010)

\bibitem{St16} Stucky, A., Scheerer, G., Ren, Z., Jaccard, D., Poumiro, J-M.,
Barreteau, C., Giannini, E., Van der Marel, D.: Scientific Reports {\bf 6}, 
37582 (2016)

\bibitem{Ed15} Edge, J.M., Kedem, Y., Aschauer, U., Spaldin, N.A., 
Balatsky, A.V.: Phys. Rev. Lett. {\bf 115}, 247002 (2015) 

\bibitem{Ke16} Kedem, Y., Zhu, J-W., Balatsky, A.V.: Phys. Rev. B {\bf
93}, 184507 (2016)

\bibitem{Pa65} Paskin, A., Singh, A.D.: Phys. Rev. {\bf 140}, A1965 
(1965)

\bibitem{Ea67} Eagles, D.M.: Phys. Rev. {\bf 164}, 489 (1967)

\bibitem{Ea15} Eagles, D.M.:  Physica B {\bf 457}, 177 (2015)

\bibitem{Fr54} Fr\"ohlich, H.: Adv. Phys. {\bf 3}, 325 (1954)

\bibitem{Ea69} Eagles, D.M.: Phys. Rev. {\bf 186}, 486 (1969)

\bibitem{Mi08} Miu, L.: Romanian Reports in Physics {\bf 60}, 717 
(2008)

\bibitem{Ka16} Kagan, M.Yu.: JETP Lett. {\bf 103}, 728 (2016)

\bibitem{Ch16} Chubukov, A.V., Eremin, I., Efremov, D.V.; Phys. Rev. B 
{\bf 93} 174516 (2016) 

\bibitem{Pe00} Perali, A., Castellani, C., di Castro, C., Grilli, M.,
Piegari, E., Varlamov, A.A.: Phys. Rev. B {\bf 62}, R9295 (2000)

\bibitem{Lu12} Lubashevsky, Y., Lahoud, E., Chaska, K., Podolsky, D., 
Kanigel, A.: Nature Phys. {\bf 8} 309 (2012) 

\bibitem{In10} Innocenti, D., Caprara, S., Poccia, N., Ricci, A., Valetta, A.,
Bianconi, A.: Phys. Rev. B {\bf 82} 174528 (2010) 

\bibitem{Ma17} Mazzioti, M.V., Valletta, A., Campi, G., Innocenti, D.,
Perali, A., Bianconi, A.: {\em arXiv}: 1705.09690 (2017)


\end{thebibliography}
\end{document}